# Frequency dependent dielectric properties of Al/ maleic anhydride (MA) /p-Si structures

S. Bilge Ocak[*a], A.B. Selçuk[b], S.B. Bayram [c] A. Ozbay[a],

[a]Gazi University, Graduate School of Natural and Applied Sciences, Gazi University, Ankara, Turkey

[b] Sarakoy Nuclear Research and Training Centre, 06983 Saray, Kazan, Ankara, Turkey

[c] Miami University, Department of Physics, Oxford, OH 45056 U.S.A.

[*]Corresponding author:

E-mail addresses; sbocak@gazi.edu.tr, semamuzo@yahoo.com

Tel/Fax: +90-312-8386800, +90-312-8385100

**Abstract:**

Al/ maleic anhydride (MA) /p-Si organic Schottky devices were fabricated on a p-Si semiconductor wafer by spin coating. The frequency and voltage dependent dielectric constant of Al/MA/p-Si have been investigated. Dielectric properties and electrical conductivity of contact structures have been investigated in detail by using spectroscopic technique in a wide range of frequencies and applied bias voltages at room temperature. The values of dielectric constant, dielectric loss, dielectric loss tangent, real and imaginary parts of the electrical modulus and ac electrical conductivity were found considerably sensitive to frequency and applied bias voltage especially in depletion and accumulation regions. Experimental results indicate that the values of dielectric constant show a steep decrease with increasing frequency for each voltage. The values of dielectric loss as a function of voltage show a jump, and dielectric loss decreases with decreasing voltage and increasing frequency. The weak increasing of the ac electrical conductivity on frequency is observed. The real part of electric modulus increases with increasing frequency. Also, the imaginary part of electric modulus shows a peak and the peak position shifts to higher frequency with increasing applied voltage. It can be concluded that the interfacial polarization can be more easily occurred at low frequencies and the majority of interface states at metal semiconductor interface contributes to deviation of dielectric properties of Al/MA/p-Si structures.

**Introduction**

The chemistry of organic molecules on silicon surfaces has attracted the attention of a great number of experimental and theoretical groups not only due to the understanding of its fundamental aspects, but also to its importance in various technological applications, such as insulator films, resist of nanolithography, chemical and biological sensors, and molecular electronics. As expected, each of these needs requires the development of a well-defined organic layer on the semiconductor surface. As such organic layers are basically formed by exposing the semiconductor surface to organic compounds, the understanding of the first stages of the interaction between the surface and the organic structure is crucial in order to improve the quality of deposited layers. The majority of the reactions between a semiconductor surface and organic molecules occur at or near the dangling bonds of the reconstructed surface. Making use of this reaction, it is possible to produce well ordered organic films on Si substrates with a stable and uniform interface. More importantly, novel organic films on silicon introduce the possibility for an entirely new physics device based on the utilization of the varieties of their functional groups. The possibility of using flexible materials for applications in the electronics and semiconductor industry has become an important issue for low-cost, large-area printed electronics, sensors, memories, radio frequency identification (RFID) tags and flexible computers [1-8].

Metal semiconductor (MS) Schottky barrier diodes with interfacial polymer such as polyaniline, poly(alkylthiophene), polypyrrole, polythiophene, poly (3-hexylthiophene), andpolyvinyl alcohol (PVA) are of great attractive research topics. Because of their potential applications and interesting properties by chemists, physicists, and electrical engineers as well [9,10]. Among the used interfacial polymer, it is the first time we used maleicanhydride (MA) which is an excellent monomer and has reactive anhydride or hydrolyzed anhydride functional groups (carboxylicgroups) [11]. MA can be polymerized by various methods [12] such as radical solution [13–15], electrochemical [16], plasma [17], UV [18] and gamma-irradiation [19], high pressure [20,21] and solid state [22] polymerizations. Low molecular weight poly (MA) is called as oligo (MA) and known as biopolymer. Poly (MA) andits derivatives are widely used in industrial cooling water, boiler water, oil-field injection, sugar mill evaporator, reverse osmosis, desalination and bioengineering applications [22–24]. However, oligo (MA) derivatives have not been studied enough. As seen in Fig. 1, the detailed information about the synthesis can be found in the article of Kahraman [12].

The electrical and dielectric properties of the MS contacts are important in the technology of semiconductors and increased by means of the selection of a proper organic semiconductor. In this respect, the interfacial parameters such as the density of interface states and the thickness of interfacial layer can influence both the electrical and dielectric behavior of these structures [25–27]. It is well known that, when localized interface states exist at the interface, the device behavior is different from the ideal case. Since the interface capacitance (excess capacitance) depend strongly on the frequency and applied voltage, the $C-V$ and $G/\omega-V$ characteristics are strongly affected [28-30]. The frequency response of the dielectric constant ($\varepsilon'$), dielectric loss ($\varepsilon''$) and dielectric loss tangent ($\tan\delta$) is dominated by a low frequency dispersion, whose physical origin has long been in question [31-33]. As far as we know, there is no work considering dependence of dielectric and electric modulus properties of Al/MA/p-Si at room temperature and applied bias voltage.

In our previous study [34], we reported the effects of electrical characteristics of the Al/MA/p-Si structures. The main purpose in this study was to achieve a better understanding of dielectric properties of Metal-Polimer-semiconductor (MPS) structures. In the present paper novel Al/MA/p-Si structures with small area were investigated at both the forward and reverse bias admittance measurements over the frequency and voltage range of 10 kHz–1 MHz and -4 to 4 V at room temperature, respectively. The variation of $\varepsilon'$, $\varepsilon''$, $\tan\delta$, $\sigma_{AC}$ and real and imaginary part of electric modulus ($M'$ and $M''$ respectively) have been investigated as a function of frequency and voltage.

## 2. Experiment

In this work, the samples were prepared on Boron doped p-type Si(1 1 1) wafer which had thickness of 280 μm and resistivity of 10 ohm. Before processing the wafer, cleaning procedures were applied. Firstly, it was dipped into acetone to remove organics for 10 min at 50°C. Then, it was washed by deionized water and released into methanol for 2 min to eliminate acetone residuals. Again, the wafer washed by deionized water and inserted in $NOH_4:H_2O:H_2O_2$ solution for 15 min at 70°C to remove oxygen from the wafer surface caused by methanol. It was dipped into deionized water to remove solution on the wafer surface. In order to take away free oxygen molecules from the surface, the wafer was bathed in 2% HF solution for 2 min. Finally, deionized water was used for cleaning procedure to complete. After cleaning, the wafer was put immediately into a vacuum chamber where Al (99.999%) was evaporated on an unpolished surface area of 1.27 cm² as ohmic contact with thickness of 640 Å with error of 2%. Then, the wafer was annealed at 500°C in vacuum for 10 min to dope aluminum into back surface of wafer. Again, the back surface of wafer was coated by Al 800 Å to complete ohmic contact. MA layer was constructed by spin coating technique. MA and dimethylformamide (DMF) were mixed in 2:1 molar ratio, and stirred for an hour. The film was deposited by spin coating at 500 rpm for 1 min and then at 1700 rpm for 45 s on polished surface of the wafer. Approximately 1 ml solution was used for coating. After spin coating the wafer, 800 Å thick Al circular rectifying contact, 1.3 mm in diameter, was deposited by evaporation at $2 \times 10^{-6}$ torr. I–V and C–V measurements were taken at room temperature to determine the electrical characteristics of the Schottky diodes. Two probe measurements were done by using a high-voltage source meter (Keithley 2410) for I–V and a precision impedance analyzer (Agilent 4294A) for C–V. Thickness of the interfacial layer was 35 nm which was evaluated from C–V measurements at 1 MHz.

## 3. Results and discussion

Studies of the frequency dependent electrical conductivity of semiconductor materials are important to explain the mechanisms of conduction in these materials. Moreover, the dielectric relaxation studies are important to understand the nature and the origin of dielectric losses, which in turn may be useful in the determination of the structure and defects in solids.

The magnitude of the dielectric constant depends on the ability of dipoles units in the material to orient with the oscillations of an external alternating electric field. The polarizable units are space charge (electronic), atomic and dipolar. Space charge polarization involves a limited transport of charge carriers (electron) until they are stopped at a potential barrier, may be due to a grain boundary or phase boundary. At lower frequency (~1 KHz) only electrons are efficiently polarized. Dipole polarization is the redistribution of charge when a group of atoms with a permanent dipole align in response to the electric field. At frequency ~$5 \times 10^4$ Hz, the dipolar effect is more active. Atomic polarization contributes to the dielectric constant by displacement of electrons in an atom relative to the nuclei. At higher frequency >$5 \times 10^6$ Hz, the atomic polarizations can occur alone. The polarizability (α) is generally additive, (i.e. α = α $_{space\ charge}$ + α$_{dipolar}$ + α$_{atomic}$) of each polarization mode to the dielectric constant (ε), (i.e. $\varepsilon^* = \varepsilon^*$ $_{space\ charge}$ + $\varepsilon^*$ $_{dipolar}$ + $\varepsilon^*$ $_{atomic}$). The complex dielectric function is expressed as [36,37]:

$$\varepsilon^* = \varepsilon' - i\varepsilon'' \qquad [1]$$

where $\varepsilon'$ and $\varepsilon''$ are the real and imaginary parts of complex permittivity, and $i$ is the imaginary root of -1. The complex permittivity formalism has been employed to describe the electrical and dielectric properties. The values of $\varepsilon'$ and $\varepsilon''$ in the frequency range from 30 kHz to 1 MHz can be calculated using [38,39]

$$\varepsilon' = \frac{C}{C_0} = \frac{C d_p}{\varepsilon_0 A}, \quad [2]$$

where $C_0$ is capacitance of an empty capacitor, $A$ is the rectifier contact area of the structure in cm², $d_p$ is the interfacial insulator layer thickness and $\varepsilon_0$ is the permittivity of free space charge. In the strong accumulation region, the maximal capacitance of the structure corresponds to the insulator capacitance ($C_{ac} = C_i = \frac{\varepsilon' \varepsilon_0 A}{d_p}$). The imaginery part of complex permittivity can be calculated using the measured conductance values from the relation:

$$\varepsilon'' = \frac{d_p G}{A \varepsilon_0 \omega}, \quad [3]$$

where $G$ is the measured conductance, and $\omega$ is the angular frequency. The dielectric loss tangent can be expressed as follows:

$$\tan \delta = \frac{\varepsilon''}{\varepsilon'}. \quad [4]$$

It is clear that $\tan \delta$ is closely related to the conductivity. The ac electrical conductivity of the dielectric material can be given by [28,40]

$$\sigma_{AC} = 2\pi f \varepsilon_0 \varepsilon' \tan \delta. \quad [5]$$

From the values of capacitance and conductance, frequency and applied voltage dependence of $\varepsilon'$, $\varepsilon''$, $\tan \delta$ and $\sigma_{AC}$ of MA/P-Si interfaces are determined in the frequency range of 30 kHz–1MHz and applied voltage range from -4 V to 4 V at room temperature. Figures 2-5 show the voltage dependencies of the values $\varepsilon'$, $\varepsilon''$, $\tan \delta$ and $\sigma_{AC}$ of MA/P-Si structures at room temperature at various frequencies. As seen from the figures, these values show a strong dependence on the applied voltage at different frequencies. It is clear that these values decrease with increasing frequency. Such behavior of $\varepsilon'$, $\varepsilon''$, $\tan \delta$ and $\sigma_{AC}$ can be explained by the fact that the interfacial dipoles have less time to orient themselves in the direction of the alternating electric field as frequency is increased [40-43].

It can be observed from Fig.2 that the values of $\varepsilon'$ is high at lower frequencies. Increasing with the decreasing frequency of dielectric constant is referred to the existence of a probable interface polarization mechanism since interface traps at low frequencies can contribute to the ac signal and the dielectric parameters. However, at high frequencies, it becomes disappeared, because the magnitude of polarization decreases with increasing frequency and the charges at states cannot follow the ac signal at high frequency. In addition, it is well known that the value of the dielectric constant depends on various parameters such as substrate temperature, annealing, frequency, the growth or preparation methods, thickness and homogeneities, applied voltage or electric field [44-48]. Also, the capacitance decreases with increasing applied frequency. This effect is believed due to the screening of the electric field across the sample by charge redistribution [25]. At low frequencies, the charges on defects are more readily redistributed, such that defects closer to the positive side of the applied field become negatively charged while the defects closer to the negative side of the field become positively charged. Similar results have been reported in the literature [25-27] and they ascribed such behavior to only interface states.

In order to explain the effect of the space charge polarization and the bias voltage, the frequency dependence of the $\varepsilon'$ and $\varepsilon''$, $\tan\delta$ of Al/ MA/P-Si at 0.5V and 1V are presented in Fig. 6a–d, As can be seen in Fig. 6 (a) and (b), the values of the $\varepsilon'$ and $\varepsilon''$ are strong functions of applied voltage and frequency. There is a sharp decrease in the values of $\varepsilon'$ and $\varepsilon''$ in the lower frequency region and it shows a frequency independent nature of the parameter in the high frequency region. The strong decrease in $\varepsilon'$ and $\varepsilon''$ with increasing frequency can be explained by means of Debye relaxation model for orientation polarization and interface effect [25]. Both $\varepsilon'$ and $\varepsilon''$ increase with increasing voltage. The dispersion in the magnitude of $\varepsilon'$ and $\varepsilon''$ at low frequencies can be referred to Maxwell-Wagner-type interfacial polarization. Since the inhomogeneities cause a conductivity which depends on frequency, charge carriers accumulate at the boundaries of less-conducting regions, so creating interfacial polarization. As can be seen in Fig. 6 (c), the values of $\tan\delta$ reveal small peaks at high frequency region and the peaks shift towards right with increasing voltages. This behavior of $\tan\delta$ can be explained by the equality of the hopping frequency of charge carriers to that of the external applied field. The value of frequency and the amount of the formed polarization determine the dielectric relaxation process [46,47].

The dielectric properties of materials can be expressed in various ways, using different representations. The terms complex impedance ($Z^*$) and complex electric modulus ($M^*$) formalisms with regard to the analysis of the dielectric or polymer materials have so far been discussed by several authors and most of them have preferred electric modulus in defining the dielectric properties and conduction mechanisms of these materials [49-51]. One of the advantages of $M^*$ formalism is that it gives more importance to the elements with the smallest capacitance occurring in the dielectric system. In the context of dielectric data, as offered by Faivre *et al.* [52], the modulus representation is occasionally used in order to emphasize small features at high frequencies. The data regarding $Z^*$ or the complex dielectric permittivity ($\varepsilon^* = 1/M^*$) can be transformed into the $M^*$ formalism using the following relation [51,52]:

$$M^* = iwC_0 Z^* \qquad [6]$$

or

$$M^* = \frac{1}{\varepsilon^*} = M' + jM'' = \frac{\varepsilon'}{\varepsilon'^2 + \varepsilon''^2} + j\frac{\varepsilon'}{\varepsilon'^2 + \varepsilon''^2} . \qquad [7]$$

The $\varepsilon^*$ and $M^*$ representation allows us to distinguish the local dielectric relation. Generally, in order to extract as much information as possible, dielectric-relation spectroscopy data are used in the electric modulus formalism introduced by Macedo *et al.* [53]. Also, the determination of the electric modulus of these materials and their variation with applied bias voltage provide valuable information that allows study of the relaxation process for a specific electronic application [53].

The real component $M'$ and the imaginary component $M''$ of the electric field are calculated from $\varepsilon'$ and $\varepsilon''$. Fig. 7 (a-b) and (c-d) show the voltage dependence of the real and the imaginary components of the electric modulus and for Al/ MA/P-Si diode at low frequencies and high frequencies at room temperature respectively. As seen in Fig. 7 (a-b) and (c-d), the values of $M'$ and $M''$ increase with increasing frequency. These behaviors are attributed to the polarization increase with increasing frequency in the Al/ MA/P-Si structures [22]. In other words, $M'$ reaches a maximum constant value corresponding to $M_\infty = 1/\varepsilon_\infty$ due to the relaxation process. At low frequencies, the values of $M'$ approach zero, confirming the removal of electrode polarization. Afterwards, the $M'$ increases exponentially with increasing voltage at all frequencies. As seen in Fig. 7(c-d), the $M''$ has a peak at all frequency and the peak shifts to lower voltages with increasing frequency, indicating that increasing frequency cause an increase in the energy of the charge carrier, thus leading to increased relaxation time [40]. The dispersion of relaxation time is attributed to the various grains and grain boundaries in the Al/ MA/p-Si device with different relaxation time [40-42]. Such behavior of one peak in $M''$ vs. V plots can be attributed to the existence of the particular density distribution profile at Al/ MA/P-Si interface. At low frequencies, almost all charges at interface states/traps can easily follow an external ac signal and they cause excess capacitance and conductance. On the other hand, the amount of tracking interface states decreases with increasing frequencies and cannot contribute to the capacitance and conductance values. The change in the dielectric parameters, electric modulus and ac electrical conductivity is a result of restructuring and reordering of charges at the interface under external electric field or voltage and interface polarization. Since the areas and capacities of these diodes are small, and interfaces states do not contribute to the values of $\varepsilon'$ and $\varepsilon''$ at high frequencies, these structures can be used in high frequency applications.

## 4. Conclusions

In summary, we have fabricated and investigated the electrical characteristics of the Al/MA/p-Si MPS Schottky structures formed by coating of the organic material to directly p-Si substrate. Frequency and voltage dependence of dielectric properties, electrical modulus and ac electrical conductivity of Al/MA/p-Si were investigated in detail in the frequency range of 30kHz – 1 MHz at room temperature. Experimental results demonstrated that the dispersion in $\varepsilon'$, $\varepsilon''$, $\tan\delta$, $M'$, $M''$ and $\sigma_{AC}$ values of diode is considerably high especially at low frequency and applied bias voltage in depletion region. The strong decrease in $\varepsilon'$ and $\varepsilon''$ with increasing frequency can be explained by means of Debye relaxation model for dipole and interfacial polarization and interface states effect. A decrease of ε′ and ε″ with increasing frequency for all operating voltages are observed. Owing to the fact that interface states do not contribute to the values of $\varepsilon'$ and $\varepsilon''$ at high frequencies and these structures can be used in high frequency applications.

**Acknowledgements**

This work is supported by Gazi University BAP office with the research project numbers 41/2012-02 and 41/2012-01. One of us (S.B. Bayram) also acknowledges financial support of the National Science Foundation (Grant No. NSF-PHY-1309571) and Miami University, College of Arts and Science.

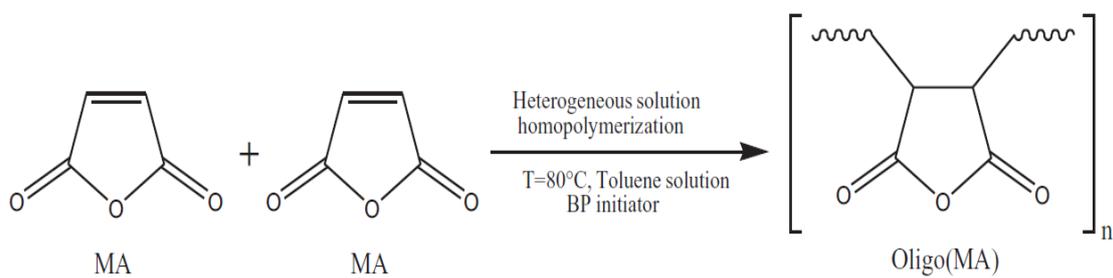

Fig. 1. Synthetic route of oliga (MA)

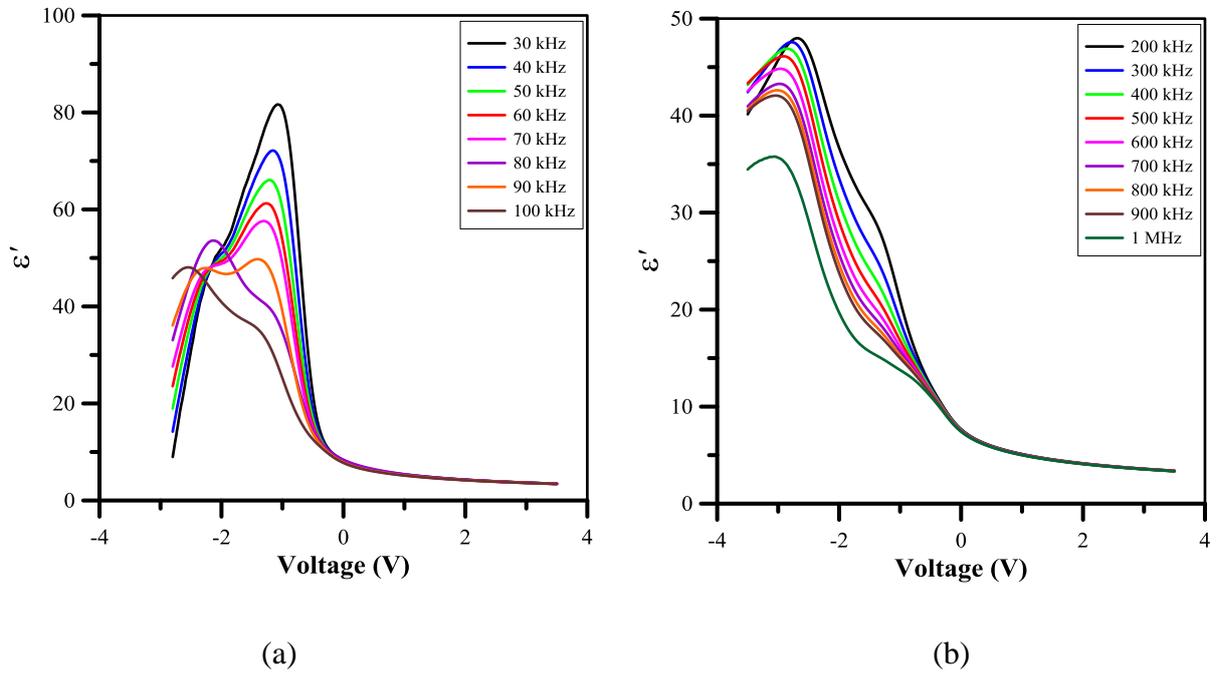

(a) (b)

Fig.2. The variation of the dielectric constant versus applied voltage of Al/MA/p-Si structure for a) low frequencies b) high frequencies at room temperature.

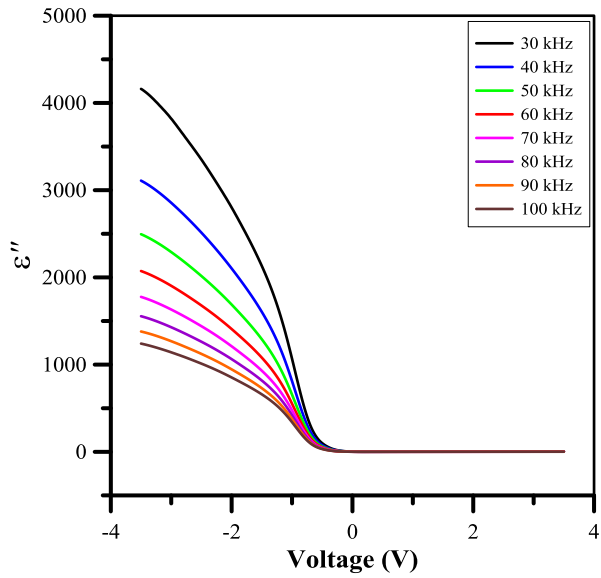 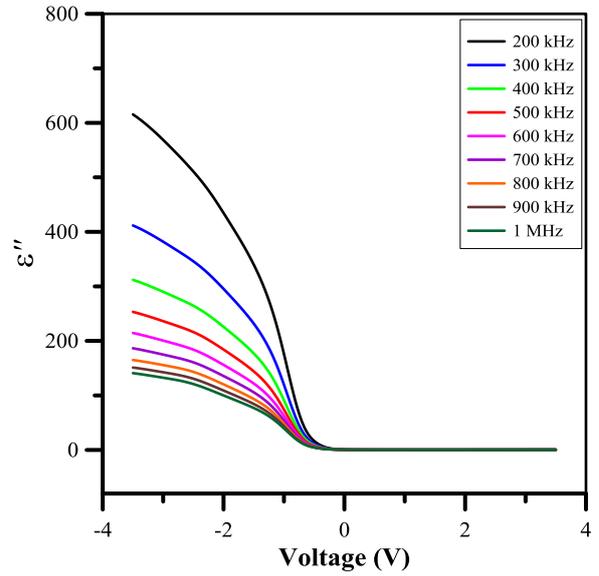

(a)                                  (b)

Fig.3 The variation of the dielectric loss versus applied voltage of Al/MA/p-Si structure for a) low frequencies b) high frequencies at room temperature.

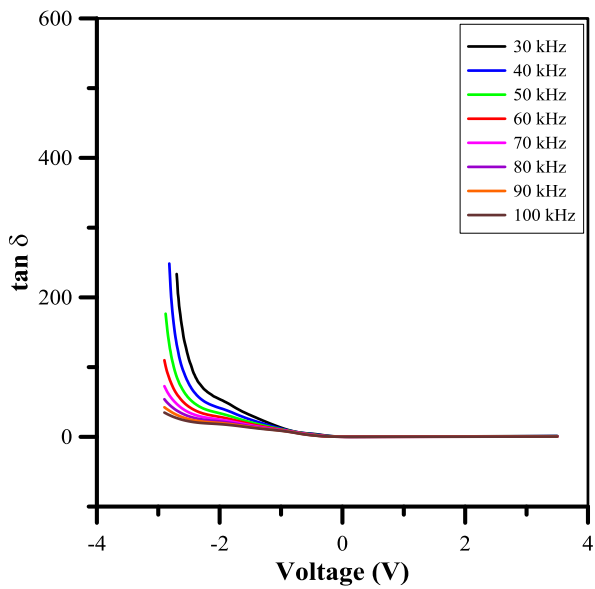 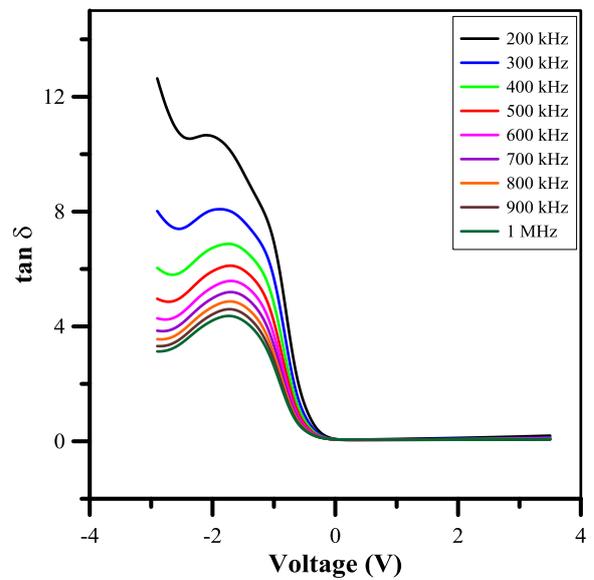

(a)                                  (b)

Fig.4 The variation of the tangent loss versus applied voltage of Al/MA/p-Si structure for a) low frequencies b) high frequencies at room temperature.

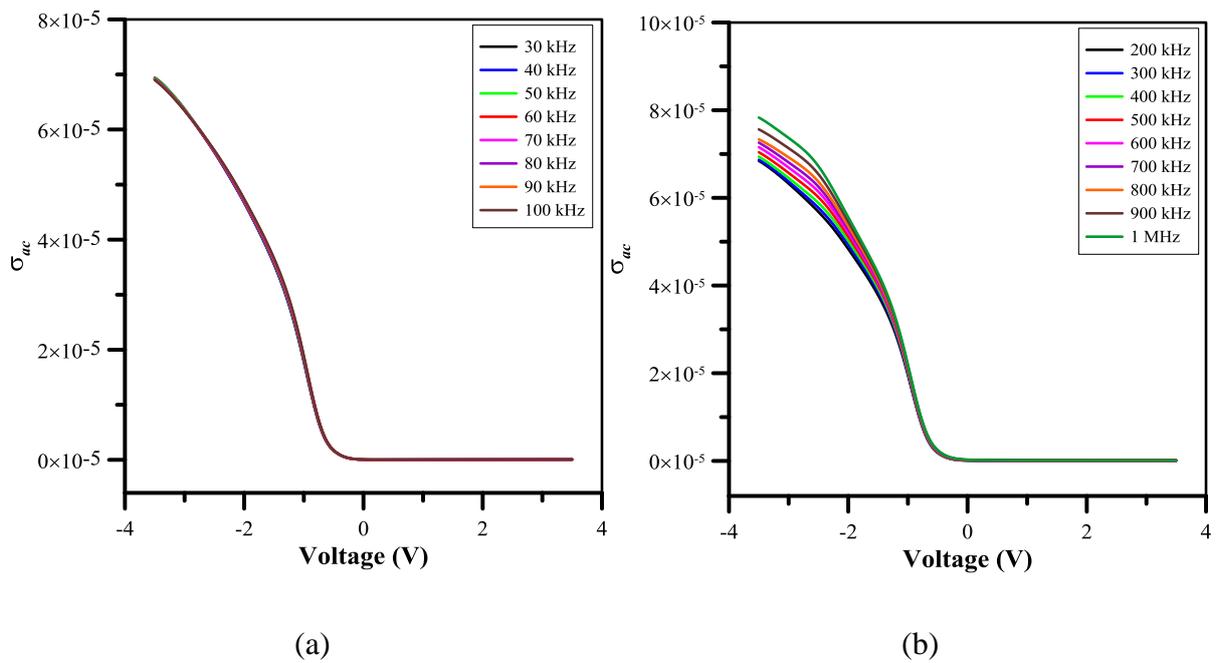

(a)          (b)

Fig.5 The variation of the AC electrical conductivity ($\sigma_{AC}$) versus applied voltage of Al/MA/p-Si structure for a) low frequencies b) high frequencies at room temperature.

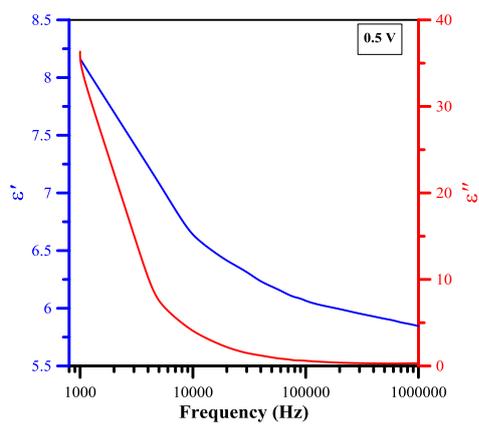

(a)

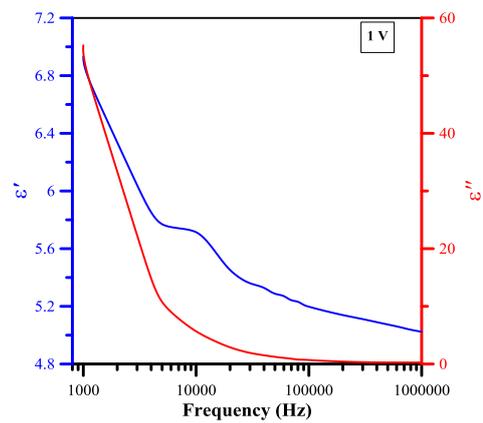

(b)

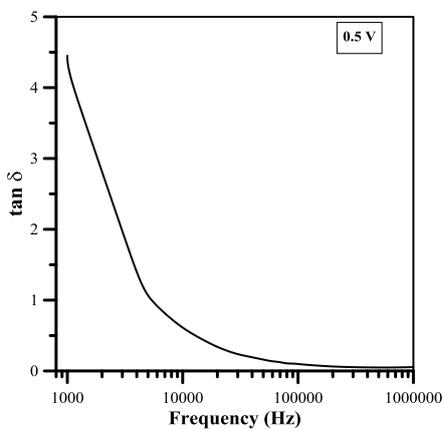

(c)

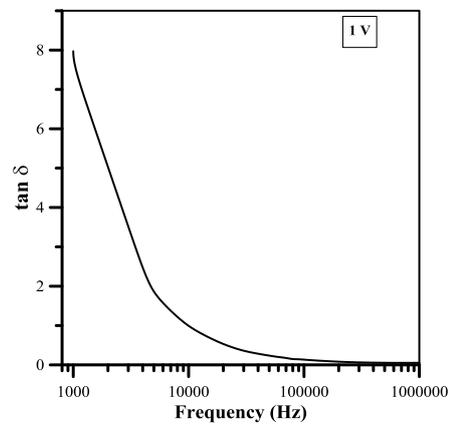

(d)

Fig.6. Frequency dependence of the dielectric constant and dielectric loss at a) -0.5V b) -1V

and frequency dependence of the tangent loss at c) 0.5V d)1V.

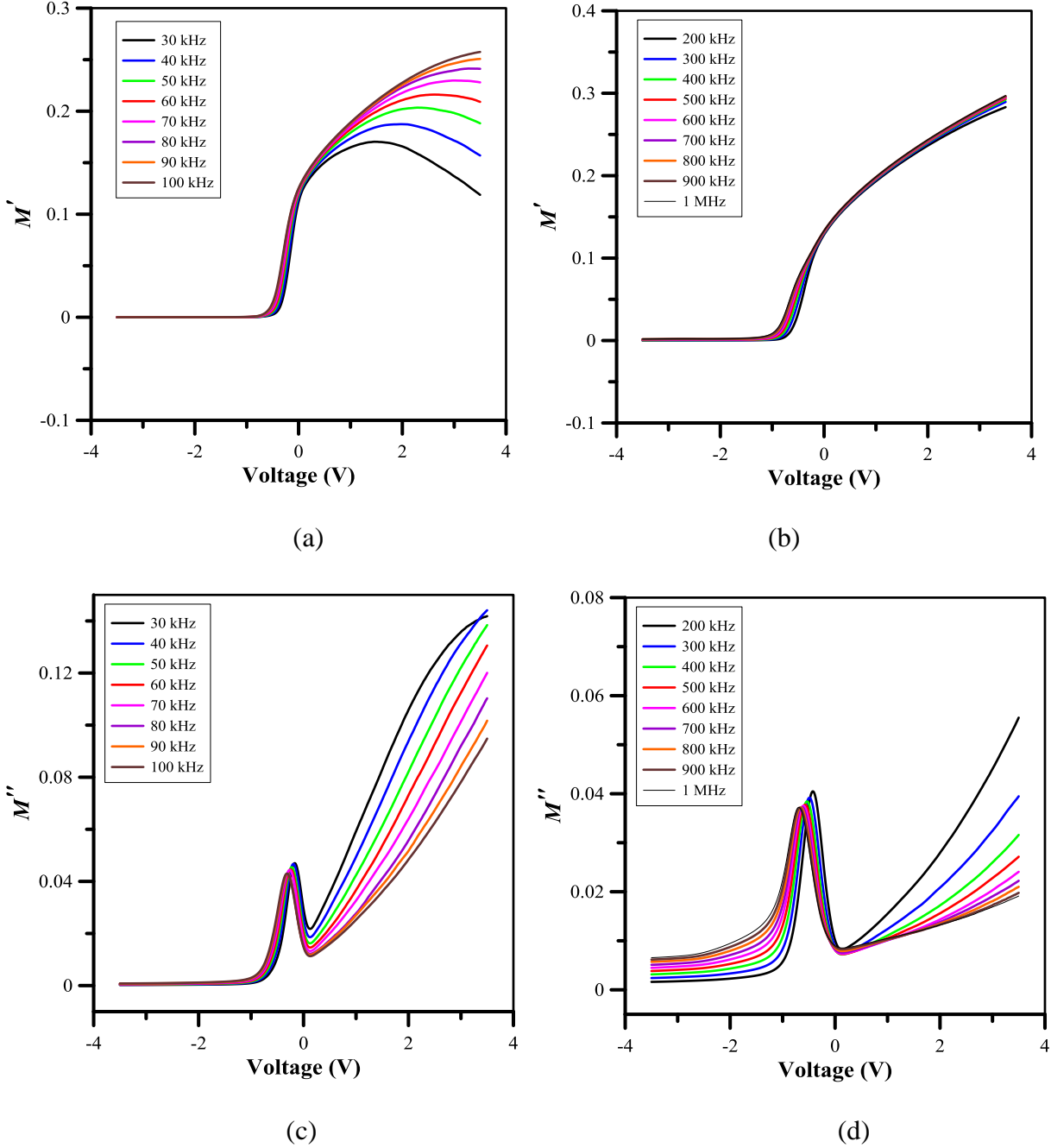

Fig.7. Variation of the $M'$ with (a) voltage between 30kHz and 100kHz. (b) voltage between 200kHz and 1MHz and variation of the $M''$ with (c) voltage between 30kHz and 100kHz (d) voltage between 200kHz and 1MHz.

17